\documentclass{aastex631}

\received{January 1, 2026}
\revised{February 1, 2026}
\accepted{March 18, 2026} 
\submitjournal{The Astrophysical Journal Letters (ApJL)}

\begin{document}

\title{Compressive Structures in the Foreshock of Collisionless Shocks}

\correspondingauthor{Savvas Raptis}
\email{savvas.raptis@jhuapl.edu/savvasraptis@pm.me}

\author[0000-0002-4381-3197]{Savvas Raptis}
\affiliation{Johns Hopkins University Applied Physics Laboratory, Laurel, MD, USA}

\author[0000-0002-0608-8897]{Domenico Trotta}
\affiliation{European Space Agency (ESA), European Space Astronomy Centre (ESAC), Camino Bajo del Castillo s/n, 28692 Villanueva de la Cañada, Madrid, Spain}

\author[0000-0002-4381-3197]{Drew L. Turner}
\affiliation{Johns Hopkins University Applied Physics Laboratory, Laurel, MD, USA}

\author[0000-0001-7171-0673]{Xóchitl Blanco-Cano}
\affiliation{Departamento de Ciencias Espaciales, Instituto de Geofísica, Universidad Nacional Autónoma de México, Mexico City, Mexico}

\author[0000-0002-3039-1255]{Heli Hietala}
\affiliation{Department of Physics and Astronomy, Queen Mary University of London, London E1 4NS, UK}

\author[0000-0003-1270-1616]{Tomas Karlsson}
\affiliation{Division of Space and Plasma Physics - KTH Royal Institute of Technology, Stockholm, Sweden}

\author[0000-0002-0606-7172]{Immanuel Christopher Jebaraj}
\affiliation{Department of Physics and Astronomy, University of Turku, 20500 Turku, Finland}

\author[0000-0002-4974-4786]{Ivan Y. Vasko}
\affiliation{William B. Hanson Center for Space Sciences, University of Texas at Dallas, Richardson, TX, USA}

\author[0000-0003-2555-5953]{Adnane Osmane}
\affiliation{Department of Physics, University of Helsinki, Helsinki, Finland}

\author[0000-0003-1434-4456]{Kazue Takahashi}
\affiliation{Johns Hopkins University Applied Physics Laboratory, Laurel, MD, USA}

\author[0000-0002-3176-8704]{David Lario}
\affiliation{NASA Goddard Space Flight Center, Greenbelt, MD, USA}

\author[0000-0002-4313-1970]{Lynn B. Wilson III}
\affiliation{NASA Goddard Space Flight Center, Greenbelt, MD, USA}

\author[0000-0003-1749-2665]{Gregory G. Howes}
\affiliation{Department of Physics and Astronomy, University of Iowa, Iowa City, Iowa 52242, USA}

\author[0000-0002-7388-173X]{Robert F. Wimmer-Schweingruber}
\affiliation{Institute of Experimental and Applied Physics, Kiel University, Leibnizstrasse 11, D-24118 Kiel, Germany}

\begin{abstract}
Collisionless shocks are fundamental accelerators of energetic particles; yet, the observations of nonlinear foreshock structures, which are essential in acceleration processes, differ significantly between Interplanetary (IP) shocks and planetary bow shocks. We present a direct comparison of two high-Mach-number, quasi-parallel shocks: an IP shock observed by Solar Orbiter and the Earth's bow shock measured by the Magnetospheric Multiscale (MMS) mission during the 2024-2025 ``string-of-pearls'' campaign. We show that Foreshock Compressive Structures (FCSs) initiate upstream of both shocks at similar normalized distances ($\lesssim$50 ion inertial lengths, $d_i$) when the suprathermal ($>10$ keV) ion density exceeds $\sim$1\% of the background. However, the IP shock lacks the fully evolved, high-amplitude Short Large Amplitude Magnetic Structures (SLAMS) characteristic of the terrestrial foreshock. We demonstrate that the ``growth zone'' capable of sustaining these structures is spatially limited ($\sim$135 $d_i$), which, due to the high speed of the propagating IP shock, corresponds to a brief observational window of $<10$ s. Beyond this observational constraint, we suggest an additional physical mechanism that can inhibit foreshock maturity at IP shocks. The lack of global curvature prevents the lateral supply (``cross-talk'') of energetic ions from different shock regions. These findings suggest that while the fundamental physics of FCS initiation is unified across collisionless shocks, the achievement of full nonlinearity can be regulated by the unique shock geometry and upstream properties, while ultimately remaining observationally challenging to identify.
\end{abstract}


\section{Introduction} \label{sec:intro}

Quasi-parallel collisionless shocks accelerate particles to relativistic energies when extended foreshock regions are sustained upstream \citep{fermi1954galactic, kennel1988shock}. These foreshocks are filled with backstreaming ions that drive instabilities, generating Ultra Low Frequency (ULF) waves \citep{gary1981electromagnetic,turc2023transmission}. As these waves advect toward the shock, they undergo nonlinear evolution, steepening into discrete Foreshock Compressional Structures (FCS). These include steepened shocklets and Short Large Amplitude Magnetic Structures (SLAMS), which are part of a broader group of intrinsic and driven foreshock transients \citep{zhang2022dayside,kajdivc2024transient}. While such transient processes are well-characterized at planetary bow shocks \citep{turner2013first,raptis2025multimission,collinson2023shocklets, zhang2025role,Wang2025SLAMS}, they remain less understood at Interplanetary (IP) shocks.

Significant physical differences between IP and Earth's bow shock exist. Fundamentally, IP shocks are locally planar and propagate as evolving forward shocks through the solar wind, whereas the bow shock is highly curved and quasi-stationary in the spacecraft's reference frame. These differences can critically influence foreshock processes. While precursor shocklets and whistlers are observed at IP shocks \citep{wilson2009low,trotta2023multi,wilson2025large}, a striking discrepancy has emerged regarding fully evolved FCS. Recent observations of strong near-Sun IP shocks reveal that, despite possessing well-developed foreshocks, they lack the high-amplitude nonlinear structures (specifically SLAMS), a typical characteristic of planetary bow shocks \citep{jebaraj2024acceleration, trotta2024properties}. This deficit has downstream consequences; given the association between FCS/SLAMS and magnetosheath jets at planetary bodies \citep[e.g.,][]{karlsson2015origin,raptis2020classifying,suni2021connection,raptis2022downstream,xirogiannopoulou2024characteristics,kramer2025jets}, the nature of IP shocks suggests potentially different formation pathways for downstream jets \citep{hietala2024candidates,osmane2024formation}.

To determine whether these environments fundamentally inhibit FCS formation or if observational factors obscure their presence, we directly compare a high Mach number IP shock observed by Solar Orbiter \citep{muller2020solar} with Earth's bow shock observed by the Magnetospheric Multiscale (MMS) mission \citep{burch2016magnetospheric}. By normalizing spatial scales and suprathermal densities, we aim to bridge the gap between these two regimes of collisionless shock physics. This comparison is enabled by instrumental advances from both missions. Solar Orbiter provides energetic particle observations at unprecedented time and energy resolutions~\citep[see, e.g.,][]{wimmer-schweingruber-etal-2021,yang-etal-2023, trotta-etal-2023,yang2025energetic}, while MMS enables high-resolution measurements of transient processes at Earth's bow shock across a wide spatial and energy scale~\citep[see, e.g.,][]{turner2021direct,lindberg2023mms,raptis2025revealing,bai2025statistical}.

\section{Data \& Methods}

We analyze an IP shock observed by Solar Orbiter on 2022-08-31 UT and an Earth bow shock crossing by MMS on 2025-02-27 UT. The MMS event utilizes the 2024--2025 ``string-of-pearls'' campaign, providing simultaneous observations of the shock ramp and the upstream environment ($>2 R_E$). A unique metric used in this study is the suprathermal ion number density ($n_{st}$), derived by combining differential flux measurements from plasma and energetic particle instruments on both missions. We integrate ion energy spectra from a cutoff of 10~keV (excluding the solar wind beam and alphas) up to the MeV range. Additionally, to facilitate direct comparison between the propagating IP shock and the quasi-stationary bow shock, we convert time-series data into a one-dimensional spatial domain normalized by the upstream ion inertial length ($d_i$). Full details regarding instrumentation modes, energy bins, calibration, and spatial transformations are provided in Appendices \ref{AppendixA} and \ref{AppendixB}.

\section{Results}

The Solar Orbiter IP shock event is a high Alfvén Mach number ($M_{A}\sim5.4$) observed on 2022-08-31 at 21:44:28 UT at 0.75 AU. For foreshock structure and shock parameter analysis, we focus on the time interval 21:00-22:00 UT. The MMS event on 2025-02-27 (02:00 - 03:00 UT) was selected due to similar Mach number ($M_{A}\sim5.8$) and shock geometry properties, and critically, because it provides the first available measurements simultaneously showing the presence and evolution of localized compressive structures from two MMS spacecraft, bounding the spatial length of their evolution due to the MMS 2024-2025 'string-of-pearls' campaign.

\begin{figure*}[ht]
    \centering
{\includegraphics[width=1\textwidth]{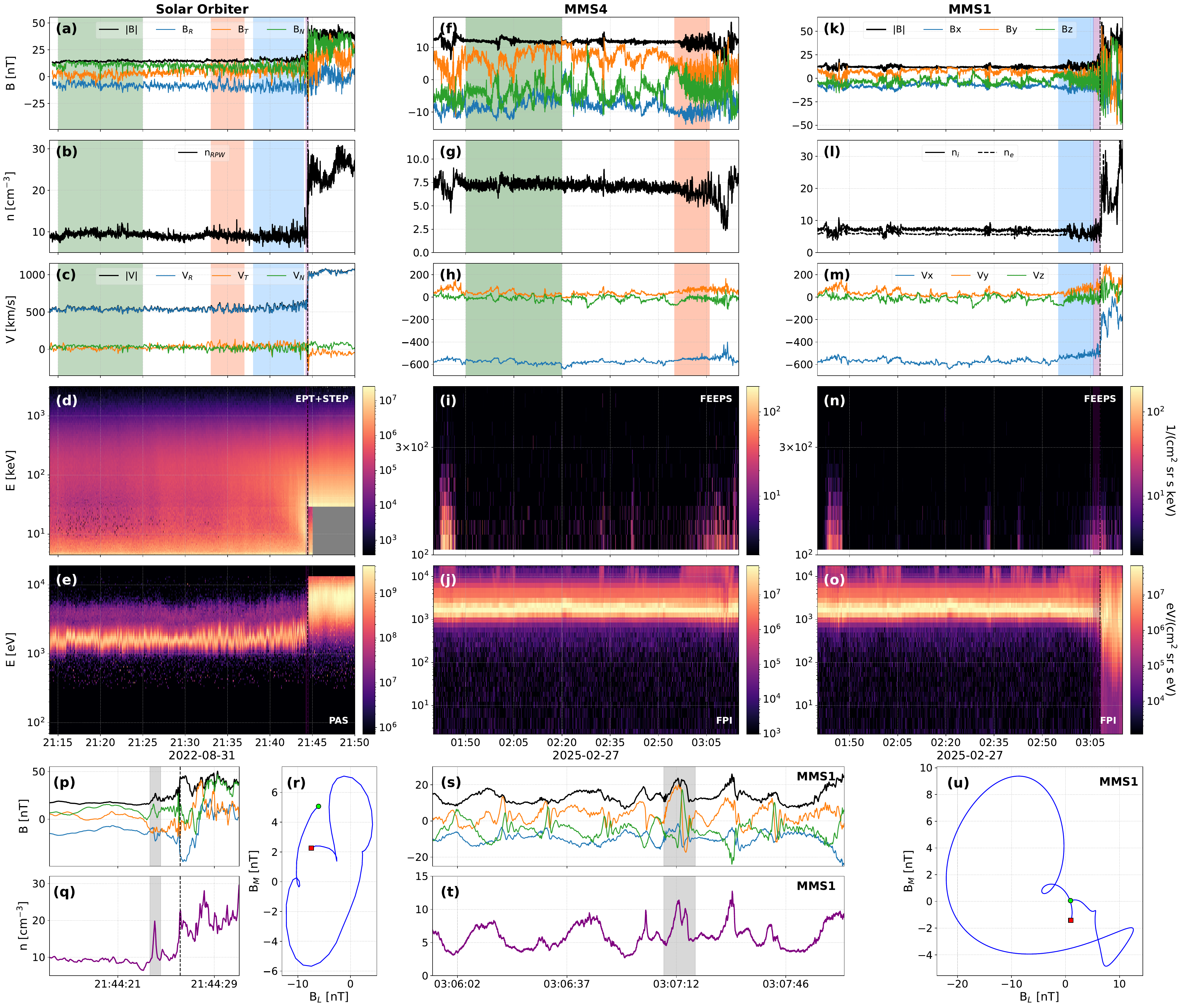}}\caption{Timeseries data for the collisionless shock observations and associated foreshock for Solar Orbiter and MMS, in RTN and GSE coordinates, respectively. Panels \textbf{(a-c, f-h, k-m)} show the magnetic field (nT), plasma density ($\text{cm}^{-3}$), and bulk ion velocity ($\text{km}\,\text{s}^{-1}$). Energetic and suprathermal ion fluxes (keV) are displayed in panels \textbf{(d, i, n)} from EPT+STEP and FEEPS, while the lower energy ion energy fluxes (eV) are shown in \textbf{(e, j, o)} from PAS and FPI. Shaded regions in the time-series panels denote different plasma environments denoted as “solar wind” (green), “far foreshock” region (orange), and “close foreshock” (blue) as described in the text and in Figure 2. The bottom row \textbf{(p-u)} details a foreshock compressive structure observed by Solar Orbiter and one of the compressive structures observed by MMS1. Panels \textbf{(p, q, s, t)} show zoomed-in time series of the magnetic field (nT) and density ($\text{cm}^{-3}$). The MVA interval is shaded in gray, while the wider interval of the zoomed-in panels is shaded in purple in the main timeseries panels. The corresponding hodograms in the L-M plane (nT) are shown in \textbf{(r, u)} with the green and red dots corresponding to start and end points, respectively. The shaded area of Solar Orbiter observations (\textbf{p,q}) and MMS (\textbf{s,t}) correspond to 1 and 10 seconds, respectively. Vertical black dashed lines denote the shock time for both the IP and the bow shock. \label{fig:1}}
\end{figure*}

To illustrate the presence of the shock, foreshock, and localized FCS, time-series plots from Solar Orbiter, MMS1, and MMS4, along with associated polarization hodograms, are presented in Figure~\ref{fig:1}. Solar Orbiter observes the IP shock crossing at approximately 2022-08-31 21:44:28 UT. For Earth's bow shock, the unique MMS constellation allows MMS4 to observe the solar wind and upstream foreshock, while MMS1 observes the foreshock environment near the shock, crossing the bow shock at approximately 2025-02-27 03:08:00 UT. Apart from shock crossings, we focus on the FCSs observed at both shocks, highlighted by gray shaded areas in the zoomed-in time-series for Solar Orbiter and MMS1 (Figure~\ref{fig:1} \textbf{p},\textbf{q},\textbf{s},\textbf{t}). At the IP shock, a well-defined FCS is observed very close to the crossing. At Earth's bow shock, MMS1 observes a series of FCS characterized by increases in density and magnetic field strength, as is typical for a quasi-parallel shock, while MMS4 observes only ULF waves with minimal steepening at the same time. Furthermore, the polarization of both highlighted structures shows an elliptical signature with durations of 1-2 seconds at the IP case and $\sim 10$ seconds at MMS (Figure~\ref{fig:1} \textbf{r},\textbf{u}). The in-phase variation of magnetic field and density indicates that the structure is a fast-magnetosonic structure, consistent with SLAMS \citep{scholer2003short}. Properties of these two collisionless shocks are summarized in Table~\ref{table:1}. While expected differences exist due to varying methodologies and instrumental intercalibration, the most important parameters are well-established: both shocks have similarly high Alfvén Mach number and a foreshock formed under quasi-parallel geometry.

\begin{table}[h]
\centering
\caption{Calculated shock parameters for the case study events in the spacecraft frame (SCF), with $M_A$ being on the shock's normal incident frame (NIF). Modeled bow shock for MMS is done using the model of \cite{merka2005three}. The range of beta parameter for MMS is sensitive to the interval used (i.e., undisturbed solar wind and foreshock), while OMNI data provide a lower estimate. See Appendix \ref{AppendixB} for additional methodological details.}
\begin{tabular}{lcc}
\hline
\textbf{Parameter [unit]} & \textbf{Solar Orbiter} & \textbf{MMS} \\
\hline
Upstream ion plasma beta (${\beta})$ & 0.3 - 0.5 & 2.0-3.0 (OMNI: $\sim$1)\\
Shock's Normal ($\hat{n})$ & [0.80, -0.52, -0.29]&  [0.939, 0.10, 0.321]\\
Modeled Shock's Normal ($\hat{n})$ &  - &  [0.88, 0.45, 0.11]\\
Shock angle ($\theta_{B\hat{n}}$) [$^\circ$] & 22 & 39 \\
Shock speed along  $\hat{n}$  ($V_{\text{sh}}$) [km/s] & 880 & 50 \\
Alfvén Mach number ($M_{\text{A}}$)& 5.4 & 5.6\\
Ion inertial length ($d_i$) [km] & 65 & 90 \\
Heliocentric distance ($L$) [AU] & 0.75 & 1.0 \\
\hline
\end{tabular}
\label{table:1}
\end{table}

\begin{figure*}[ht]
    \centering
{\includegraphics[width=1\textwidth]{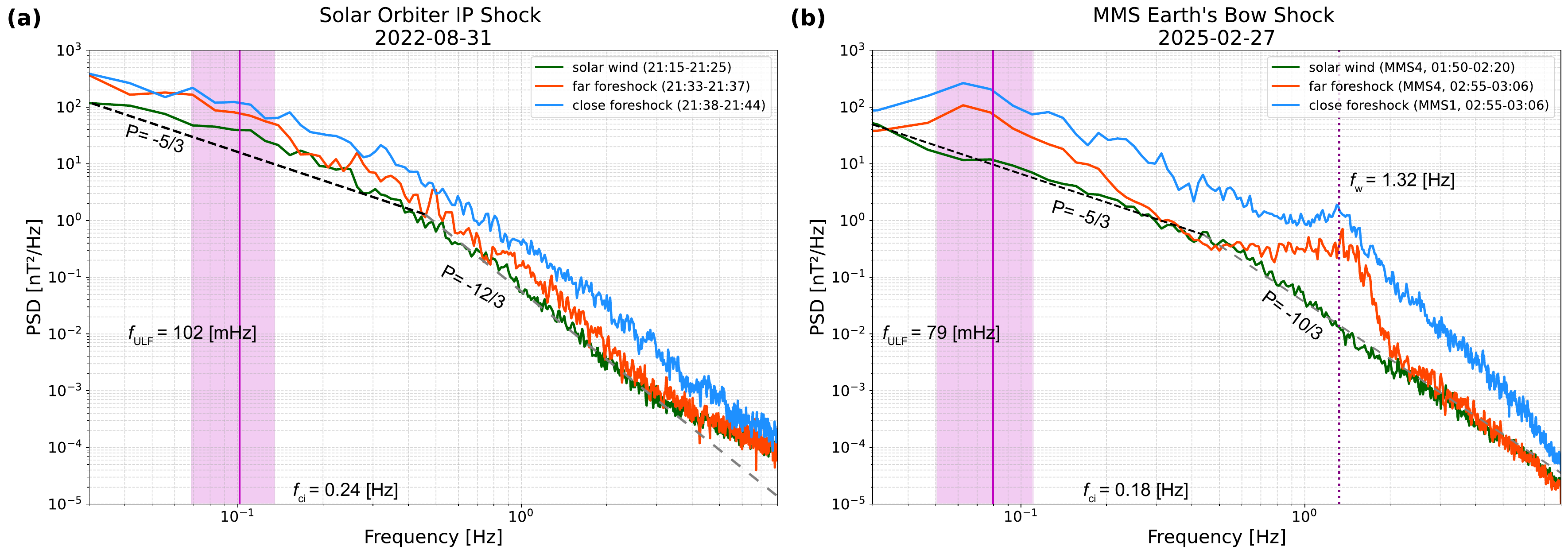}}\caption{A comparison of magnetic field power spectral densities (PSDs) in the foreshock of the IP shock and Earth's bow shock in the spacecraft frame. For both panels, power-law slopes for different turbulent regimes are shown for reference, with stronger dissipation observed at the IP shock. Panel \textbf{(a)} shows PSD of the magnetic field from the Solar Orbiter spacecraft on August 31, 2022, upstream of the shock. Three distinct plasma environments are compared: the “close foreshock” immediately upstream of the shock (blue), a “far foreshock” region (orange), and a “solar wind” upstream interval representing the background solar wind (green). Characteristic frequencies, including the local ion cyclotron frequency ($f_{\text{ci}}$) and the expected ULF wave frequency \citep{takahashi1984dependence}, are indicated in the plot. \textbf{(b)} shows a similar analysis for Earth's bow shock on February 27, 2025, observed by the MMS mission. The plot compares the “close foreshock” (MMS1, blue) with the “far foreshock” (MMS4, orange) and a quiet upstream solar wind interval (MMS4, green). In addition to the $f_{\text{ci}}$ and expected frequencies, the approximate location of the secondary peak associated with the presence of whistler waves is shown ($f_w \approx 1$~Hz). The shaded area for the expected ULF range is obtained by assuming an error of $\pm$ 3 nT in upstream magnetic field and $\pm10^\circ$ degrees in the determination of $\theta_{\text{Bn}}$. The intervals used per line are also visualized as shaded areas in Figure 1. \label{fig:2}}
\end{figure*}

We evaluate the foreshock wavefield of both shocks by examining the power spectral density (PSD) of the magnetic field upstream of each shock. As shown in Figure~\ref{fig:2}, the background solar wind (green line) exhibits a turbulent magnetic energy spectrum consistent with expected slopes ranging from the inertial to the sub-ion kinetic range, with a break frequency at approximately $f \sim 0.4$~Hz, aligning with typical observations near Earth and close to the Sun \citep{howes2008model,bruno2013solar,kiyani2015dissipation,howes2015dynamical}. However, upon approaching the foreshock, such turbulence-related slopes do not describe the field dynamics, as the field becomes dominated by local kinetic processes rather than the ambient turbulence cascade. Focusing on foreshock effects, we note key similarities and differences. A primary similarity is a relative increase in power in the ULF range when moving closer to the shock (Green $\rightarrow$ Orange $\rightarrow$ Blue line; see shaded areas in the top panels of Figure~\ref{fig:1}), consistent with empirical formulas from Earth-based observations \citep{takahashi1984dependence}. However, there are distinct differences. Solar Orbiter observations (Figure~\ref{fig:2}\textbf{a}) show a more broadband increase in PSD as we move closer to the shock, notably lacking any distinct peak above the ion cyclotron frequency ($f_{\text{ci}}$). In contrast, MMS observations exhibit a clear secondary peak forming at an approximate frequency corresponding to low frequency whistler (fast magnetosonic) waves ($f_{\text{w}}\approx 1$ Hz) \citep{hoppe1980whistler,hoppe1981upstream}. This peak becomes more prominent closer to the shock (MMS1) and may be partially attributed to dispersive whistler waves originating from foreshock compressive structures. Such waves are also observed at the foot of collisionless shocks \citep{wilson2017revisiting,lalti2022whistler,balikhin2023structure,amano2024statistical,WimmerSchweingruber2025}, and due to similar dynamics, can be found upstream of localized shock-like structures, such as shocklets and SLAMS \citep{wilson2013shocklets,raptis2022downstream,wang2024statistical,Wang2025SLAMS}.

\begin{figure*}[ht]
    \centering
{\includegraphics[width=1\textwidth]{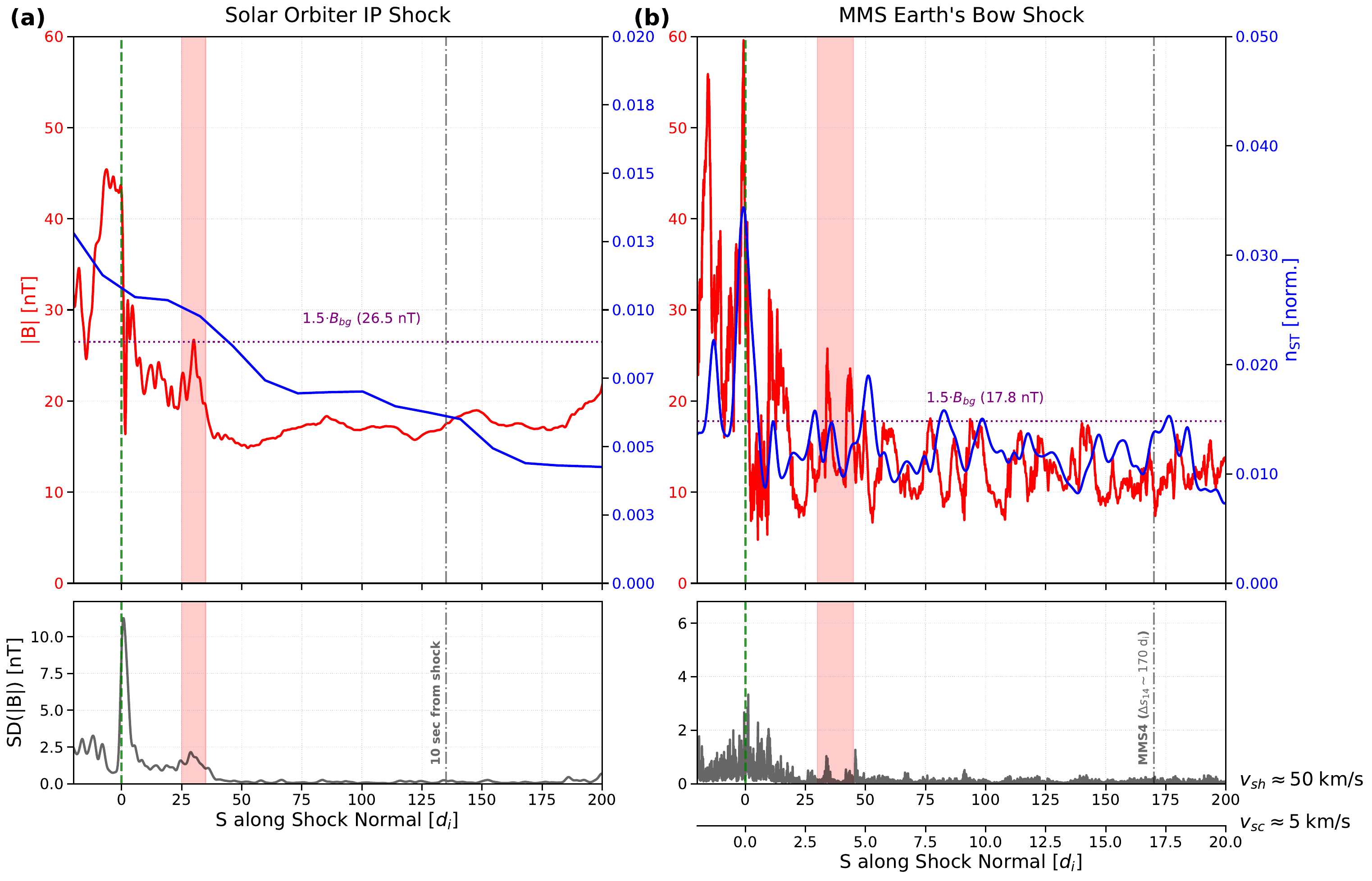}}\caption{Comparative analysis of magnetic field and suprathermal particle properties across collisionless shocks. Panel\textbf{ (a)} shows Solar Orbiter IP shock observations, while panel \textbf{(b)} presents MMS1 measurements of Earth's bow shock. For each shock event, the upper plots display the magnetic field magnitude $|B|$ (red line) and normalized suprathermal particle density ($n_{st}/n_{sw}$; blue line) as functions of distance along the shock normal. The lower plots show the magnetic field standard deviation $\sigma$(B) using a 15-data point moving window, which quantifies local variability levels. The x-axis represents distance in upstream ion inertial lengths ($d_i$) from the shock crossing ($S = 0$). The background magnetic field is defined as the average value from observations taken at $>$100 $d_i$. Red shaded regions highlight the presence of “compressive structures”  occurring at the same relative distance of $\sim 25-50$ $d_i$. The two horizontal axes on MMS plots indicate the spatial distance computed with the calculated shock speed along the normal and with the spacecraft speed, representing the upper and lower bounds of the distance from the shock, respectively, while the relative location of MMS4 bounds this in context to Figure \ref{fig:1}. More information and details about the suprathermal density and their profiles are detailed in Appendix \ref{AppendixB} and in the associated supplementary plots Figures \ref{fig:appendix_1} and Figures \ref{fig:appendix_2}. \label{fig:3}}
\end{figure*}

We continue our analysis by focusing on the compressive structures observed upstream of both shocks. This comparison is strongly influenced by different observational geometries and spatiotemporal scales involved. The propagating IP shock plows through the solar wind, essentially providing a rapid one-dimensional spatial cut of the foreshock region as it passes the spacecraft. In contrast, Earth's quasi-stationary bow shock allows MMS to observe the temporal evolution within a relatively static foreshock, filled with scattered particles. Therefore, to analyze events on similar baseline, we transform the temporal profiles (Figure~\ref{fig:1}) into spatial evolution relative to the shock. Figure~\ref{fig:3} shows this transformation by plotting three key quantities: magnetic field Standard Deviation (SD), magnetic field magnitude, and suprathermal density normalized to background solar wind density versus ion inertial lengths as distance from the shock (see Appendix~\ref{AppendixB}). In contrast to the temporal picture in Figure~\ref{fig:1}, this spatial mapping reveals that compressive structures are observed at $\lesssim 50~d_i$ from the shock in both cases. Furthermore, the increase in suprathermal density in both cases correlates well with proximity to the shock. In terms of differences, the normalized suprathermal density of the IP shock is lower than that of the bow shock (note the different values on the right vertical axes). However, there are major differences between instruments. MMS is expected to overestimate the suprathermal density ratio, while Solar Orbiter can underestimate it (Appendix B). Therefore, while physical factors may contribute to this difference, our primary explanation is that these differences are mostly driven by instrumental effects. Nevertheless, from this analysis, it appears that in high Mach number shocks, about $\sim 1\%$ of suprathermal ions (as estimated from the $n_{st}/n_{sw}$ ratio shown in Figure~\ref{fig:3}) enable nonlinear feedback and the formation of localized compressive structures to be observed at both IP and bow shocks. Furthermore, examining this in conjunction with the analysis shown in the Appendices (Figures~\ref{fig:appendix_1} and \ref{fig:appendix_2}), we find that the effective foreshock region of the IP shock that can be directly compared to Earth's observations is constrained to $\sim 135~d_i$. Finally, by using the MMS1-MMS4 separation ($\sim 170~d_i$) as a spatial bound for the Earth's foreshock region, it becomes evident that the equivalent foreshock observations correspond to a very narrow $\sim 10$-second interval of the IP shock timeseries.

\section{Summary \& Discussion}
Our results can be summarized as follows:

\begin{enumerate} 
    \item We present observational evidence of localized foreshock compressive structures (FCS) at an IP shock with a duration shorter than the local ion cyclotron period. This differs from SLAMS at Earth presented here and in other statistical works \citep{bergman2025statistical}. While not fully evolved, as they do not reach as high magnetic field magnitude amplitudes (See Figures~\ref{fig:1} and \ref{fig:3}), these structures appear localized, caused by the intrinsic wave field of the foreshock in proximity to the shock, and correspond to elliptically polarized structures in the SC-frame, observed between the shock ramp and $\lesssim 50~d_i$.

  \item The foreshock regions of both shocks display elevated wave activity and upstream suprathermal ions. However, unlike the clear spectral signatures seen in the MMS case, whistler waves associated with the Solar Orbiter shock are faint. Apart from physical reasons such as the absence of well-developed foreshock compressive structures and the rapid transit of the shock structure relative to the spacecraft, such waves may be limited in amplitude by magnetic field gradients or may be sufficiently Doppler-shifted to approach the observational limits of the Solar Orbiter instrument.

    \item After calibrating Solar Orbiter observations, upstream suprathermal ion density ($>10$~keV) is similar between the two shocks (Figures~\ref{fig:3} and \ref{fig:appendix_1}). When normalized with respect to solar wind bulk density ($n_{st}/n_{sw}$), the presence of compressive structures appears at about $\gtrsim1\%$ suprathermal density near the shock.
  
    \item The effective foreshock region of the IP shock that can be compared to Earth's observation is constrained to $\sim 135~d_i$, corresponding to only $\sim 10$ seconds of the time series. This spatial scale is bounded by the MMS1-MMS4 separation ($\sim 170~d_i$) (See Figures~\ref{fig:3}, \ref{fig:appendix_1}, and \ref{fig:appendix_2}). The ULF waves observed by MMS4 exhibit minimal non-linear evolution, thereby constraining the spatial scale required for SLAMS formation.

    \item IP shocks generate spatially extended foreshock regions due to high-energy particles ($>50$ keV) accelerated across the vast scales of their geometry. However, for the near-shock upstream region ($\lesssim 150~d_i$), the suprathermal ion density ($> 10$ keV) contributed by these energetic particles remains insufficient to drive significant wave growth and subsequent nonlinear evolution into mature SLAMS (See Figures~\ref{fig:3}, \ref{fig:appendix_1}, and \ref{fig:appendix_2}).
    
\end{enumerate}

The comparison between Solar Orbiter and MMS observations reveals that while localized foreshock compressive structures (FCS) are initiated at quasi-parallel IP shocks, their development into fully evolved, high-amplitude SLAMS is significantly inhibited compared to the terrestrial bow shock. Although differences in upstream ion plasma $\beta$ (Table~\ref{table:1}) may influence growth rates, the comparable normalized suprathermal ion densities suggest that the scarcity of mature structures at the IP shock is likely driven by differences in the shock's geometry and by observational capabilities.

The curvature of the terrestrial bow shock enables ``cross-talk'' between different shock geometries. Ions accelerated in adjacent shock regions can drift along the shock face and populate the foreshock, providing a consistent and diverse seed population that maintains a steady-state, fully developed foreshock environment. Conversely, while IP shocks are not strictly planar and exhibit localized rippling that can drive bursty acceleration \citep[e.g.,][]{trotta-etal-2023}, they are effectively planar on the scales of foreshock formation observed here. This lack of global curvature eliminates the lateral supply of energetic ions into the foreshock region. While IP shocks do generate spatially extended foreshock regions from high-energy particles ($>$50 keV) \citep{jebaraj2024acceleration,berland2025parker} accelerated across the vast scales of their geometry, at the near-shock upstream region ($\lesssim150~d_i$), the suprathermal ion density ($>$10 keV) contributed by these energetic particles appears to be insufficient to drive significant wave growth and subsequent nonlinear evolution into mature SLAMS. 

\leavevmode Furthermore, we must account for observational limitations imposed by the different environments. As detailed in Appendix~\ref{AppendixB}, the spatial transition of the suprathermal population occurs over $\sim$135~$d_i$. At Earth's bow shock, a spacecraft might dwell within such a region for minutes to hours, allowing for characterization of spatial scales and growth. However, due to the rapid motion of the IP shock, this critical evolutionary zone passes Solar Orbiter in $\sim$10 seconds. This implies that the scarcity of observed SLAMS-like structures in IP shock literature may not solely be due to physical inhibition, but also a probabilistic result of the ``growth zone'' being spatially limited and temporally short-lived. The structures observed here ($\lesssim$50$~d_i$) possibly represent the initial stages of SLAMS that are overtaken by the shock before they can achieve the nonlinear amplitudes characteristic of the terrestrial foreshock. However, a factor that warrants further investigation is whether kinematic differences between the propagating IP shock and the quasi-stationary terrestrial bow shock restrict the dwell time of foreshock waves within the growth zone. One can hypothesize that, depending on the group velocity of the generated waves relative to the shock frame, a wave packet may have less time to evolve non-linearly at an IP shock compared to the bow shock. Verifying this hypothesis requires a more rigorous analysis of wave propagation versus convection timescales and is highly dependent on the specific nature of the instability generating the waves. This remains an open question for future study.

Finally, an additional avenue warranting future investigation concerns the role of the ambient solar wind conditions set by the heliocentric distance of each shock. Solar Orbiter observed the IP shock at $\sim$0.75~AU, while MMS orbits Earth at $\sim$1~AU (Table~\ref{table:1}). Beyond the shock parameter differences already discussed, the background solar wind itself evolves with heliocentric distance \citep[e.g.,][]{liu2024radial}, with both the solar wind magnetization and the intensity of pre-existing turbulence varying significantly. The latter may have an effect on foreshock compressive structures by modulating the efficiency of ion reflection and suprathermal particle acceleration at the shock \citep{subashchandar2025parallel}. As shown in Figure~\ref{fig:2}, the two shocks studied here indeed exhibit slightly different upstream turbulence profiles prior to each shock encounter. Whether such differences can contribute to foreshock dynamics, for example, by altering the seed population available for wave excitation or by modifying the nonlinear growth rates of compressive structures, remains an open and promising question for future comparative studies.

\section{Conclusion}
Future investigations offer several promising avenues to generalize these findings. A primary approach involves performing kinetic simulations with varying shock scales, geometries, and upstream ion plasma $\beta$ to better isolate the physical mechanisms driving foreshock evolution. Observational follow-up studies should unfold the relationships between distance from the shock and suprathermal densities (see Figures~\ref{fig:3}, \ref{fig:appendix_1}, and \ref{fig:appendix_2}) by analyzing individual energy bins across all instruments. Furthermore, a systematic, multi-spacecraft analysis utilizing the MMS ``string-of-pearls'' campaign observations is needed to quantify the rapid spatial evolution of the foreshock wavefield and its compressive structures. While a broader statistical study remains challenging due to the rarity of high-Mach-number, quasi-parallel IP shocks, the inclusion of Parker Solar Probe \citep{raouafi2023parker} measurements may partially alleviate this issue.

However, a critical outcome of our analysis is the identification of a fundamental spatio-temporal disparity that must guide such future efforts. We find that the mature foreshock environments captured over hours of observations by planetary missions like MMS can correspond to a brief interval of only $\sim10$ seconds near IP shocks. This drastic difference implies that the reported scarcity of foreshock structures at IP shocks can be a consequence of both the physical limitations and the observational constraints of insufficient sampling duration. Consequently, future comparative analyses must interpret results with extreme caution, recognizing that these fundamental observational biases are further complicated by distinct interaction geometries.

Ultimately, despite these challenges, this comparative study shows that IP shocks behave largely similarly to planetary bow shocks, as expected from collisionless shock theory. While statistical analysis is necessary, this case study provides evidence that quasi-parallel IP shocks exhibit typical ion foreshock properties and wave evolution, including localized formation of compressive structures. Although environment specific properties such as the lack of geometric ``cross-talk'' may limit the full maturity of the foreshock environment, the fundamental physics governing wave generation and nonlinear steepening at IP and planetary shocks appears unified.

\section*{Acknowledgments}

\begin{acknowledgments}
SR acknowledges funding from the MMS Early Career Award 80NSSC25K7353 and the Magnetospheric Multiscale (MMS) mission of NASA's Science Directorate Heliophysics Division via subcontract to the Southwest Research Institute (NNG04EB99C). SR also acknowledges the support of the International Space Sciences Institute (ISSI) team 555, “Impact of Upstream Mesoscale Transients on the Near-Earth Environment”, and the useful discussions with Ahmad Lalti.  XBC thanks PAPIIT DGAPA  IN106724 and SECIHTI CBF-2023-2024-852 grants. H.H is supported by the Royal Society Award URF\textbackslash R1\textbackslash180671. URF\textbackslash R \textbackslash251031. I.C.J. acknowledges support from the Research Council of Finland (X-Scale, grant No.~371569). K.T. was supported by the Parker Solar Probe project under contract NNN06AA01C. R.F.W-S. acknowledges support of Solar Orbiter's EPD by DLR grant 50OT2002. We also acknowledge support from ESA through the Science Faculty - Funding reference ESA-SCI-E-LE-170. \\

Magnetospheric Multiscale (MMS) data can be found through the public SDC \url{https://lasp.colorado.edu/mms/sdc/public/about/browse-wrapper/} or the Graphical User Interface (GUI) found in \url{https://lasp.colorado.edu/mms/sdc/public/search/}.  Solar Orbiter measurements are accessible via \url{https://soar.esac.esa.int/soar/}.
\end{acknowledgments}

\vspace{5mm}

\software{Python 3.12 with typical packages and Pyspedas \citep{grimes2022space} was used for partially downloading and preprocessing data and for the figures associated with Solar Orbiter observations. MATLAB 2024b with IRFU-Matlab \citep{khotyaintsev_2024_14525047} (v1.17.0) was used for accessing the data from MMS. The PSD spectrum was generated using \texttt{scipy.signal.welch} (v1.13.0), while the smoothing of magnetic field data for illustrative purposes was done via \texttt{scipy.ndimage.gaussian\_filter1d}(v1.13).}

\appendix

\section{Instrumentation Details}\label{AppendixA}

\subsection{Solar Orbiter}
Solar Orbiter observational capabilities were fully utilized for the 2022-08-31 event. Magnetic field measurements are obtained from the flux-gate magnetometer~\citep[MAG;][]{Horbury2020}, operating at 64 vectors/s in burst mode. Ion energy flux and ground moments are provided by the Proton Alpha Sensor (PAS) of the Solar Wind Analyzer (SWA) suite~\citep[][]{Owen2020} at 4s resolution. Solar wind electron density is estimated from the spacecraft potential using the Radio and Plasma Waves instrument~\citep[RPW;][]{Maksimovic2020} at a high time resolution of 0.01~s. 

Energetic particles are measured by two instruments of the Energetic Particle Detector (EPD) suite~\citep{RodriguezPacheco2020}. We utilize the SupraThermal Electrons and Protons (STEP) sensor, which measures ions (and electrons) from a few keV to $\sim$80~keV using 15 sunward-pointing pixels. We also use the sunward aperture of the Electron Proton Telescope (EPT), measuring ions from 25~keV to 6.4~MeV (and electrons to 0.475~MeV). Both EPT datasets are used at 1-second resolution. The fields of view of these instruments overlap and are approximately 30 degrees wide. All Solar Orbiter vectors are presented in spacecraft-centered Radial-Tangential-Normal (RTN) coordinates~\citep{Franz2002}, where R points radially outward from the Sun towards the spacecraft, T is roughly along the orbital direction, and N is northward with the RN plane being on the solar rotation axis.

\subsection{Magnetospheric Multiscale (MMS)}
MMS measurements are derived from the 2024--2025 campaign. Observations primarily use Level 2 data from the Fast Plasma Investigation \citep[FPI;][]{pollock2016fast}, providing ion and electron plasma moments at 4.5~s (survey) and partially 30~ms (burst) resolution. Regarding solar wind density, we use the electron moments provided by FPI. Although spacecraft potential techniques can improve density accuracy, they are not offered as a readily available product; therefore, the next optimal moment is the one from the electron distribution of FPI \citep{roberts2021study}. 

Complementary ion measurements are obtained from the Fly's Eye Energetic Particle Spectrometer \citep[FEEPS;][]{blake2016fly} at 2.5~s resolution. To assess suprathermal ions, we use the Energetic Ion Spectrometer \citep[EIS;][]{mauk2016energetic}, measuring protons ($\sim$20~keV), helium ($\sim$60~keV), and oxygen ($\sim$130~keV) up to $\sim$1~MeV. EIS provides complete angular distributions within one spacecraft spin ($\sim$20~s). Magnetic field data are taken from the Fluxgate Magnetometer \citep[FGM;][]{russell2016magnetospheric} operating at 16~Hz. All MMS vector quantities are expressed in Geocentric Solar Ecliptic (GSE) coordinates. MMS1 and MMS4 were separated by $>2 R_E$, with MMS1 located at (12.16, -5.68, 1.32 $R_{\text{E}}$) and MMS4 at (14.85, -5.78, 0.72 $R_{\text{E}}$). OMNIweb solar wind data \citep{king2005solar} were used for validation.

\section{Methodological Details}\label{AppendixB}

\subsection{Suprathermal Density Derivation}
The suprathermal ion number density ($n_{st}$) is derived by combining differential flux measurements from distinct instruments:
\begin{equation}
n_{st} = 4\pi \int_{E_{min}}^{E_{max}} \frac{j(E)}{v} \, dE
\end{equation}
where $v$ is the ion velocity and the lower limit is set to $E_{min}=10$~keV to exclude the core solar wind beam.

\begin{figure*}[ht!]
    \centering
    \includegraphics[width=\textwidth]{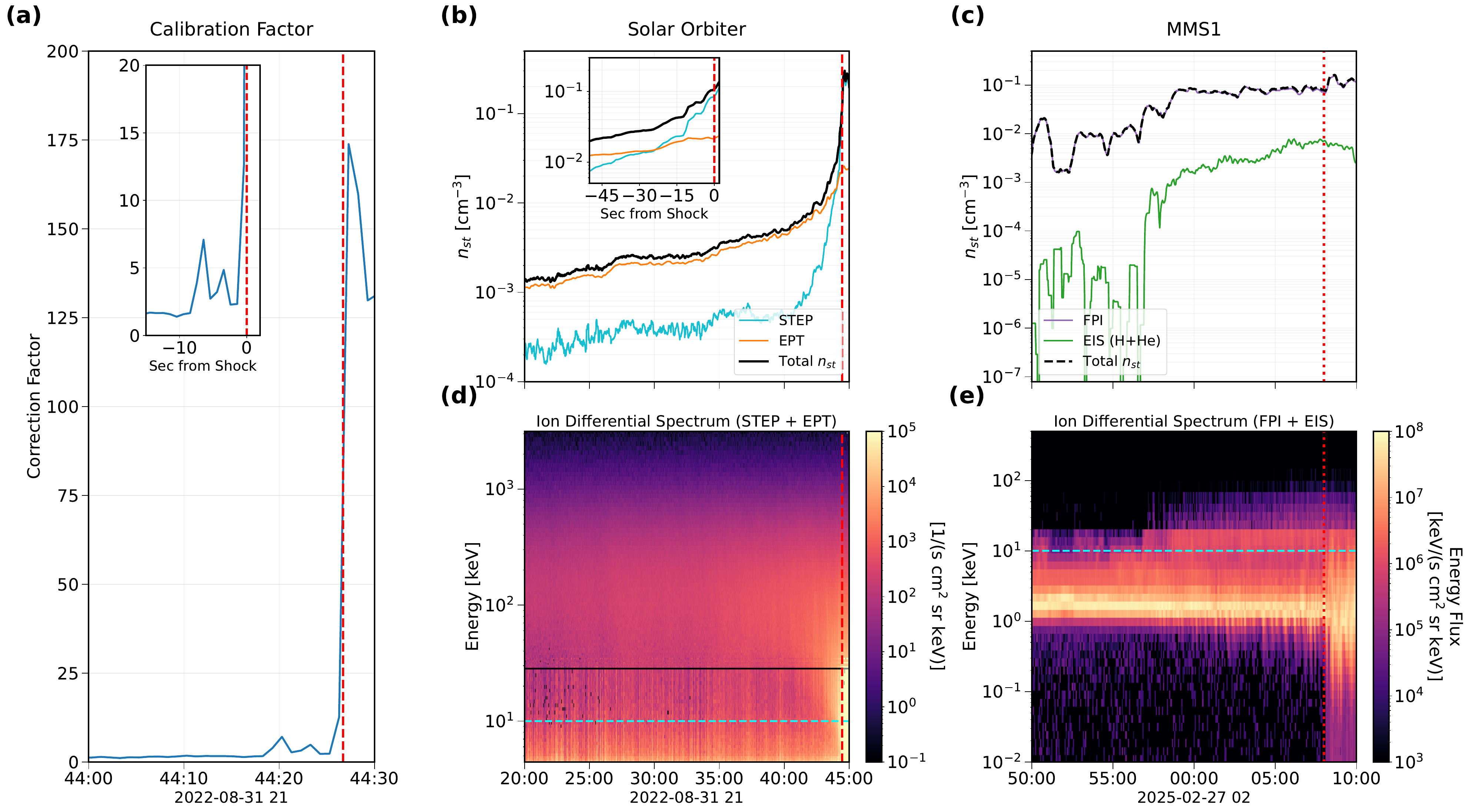}
    \caption{Derivation of suprathermal ion density ($n_{st}$) for Solar Orbiter and MMS.
    \textbf{(a)} Time-varying calibration factor for Solar Orbiter STEP data, derived to correct for dead-time saturation near the shock (red dashed line).
    \textbf{(b)} Solar Orbiter $n_{st}$ time series. The density (black) sums the STEP (cyan) and EPT (orange) components. The inset highlights the rapid density increase driven by STEP immediately upstream.
    \textbf{(c)} MMS $n_{st}$ time series, combining contributions from FPI (purple, $<$30 keV) and EIS (green, $>$30 keV). The vertical red dotted line marks the shock crossing.
    \textbf{(d)} Combined Solar Orbiter dynamic spectrum (STEP + EPT) used for integration.
    \textbf{(e)} Combined MMS dynamic spectrum (FPI + EIS). Cyan dashed lines on panels \textbf{(d)} and \textbf{(e)} show the minimum energy used for the suprathermal density integration}
    \label{fig:appendix_1}
\end{figure*}

For MMS, $n_{st}$ combines FPI and EIS data. EIS data is linearly interpolated to the FPI cadence. A composite spectrum is constructed (Figure~\ref{fig:appendix_1}e), integrating FPI fluxes from the cutoff up to $E_{max,FPI}$ ($\sim$30~keV), supplemented by EIS fluxes for the high-energy tail. FPI represents the main contribution of suprathermal density ($n_{\text{FPI}}$), with EIS ($n_{\text{EIS,p}} + n_{\text{EIS,He}}$) contributing only $\sim 1\%$ to the total. We note that using FEEPS instead of EIS produces a negligible difference ($<1\%$).

\begin{figure}[ht!]
    \centering
    \includegraphics[width=0.8\textwidth]{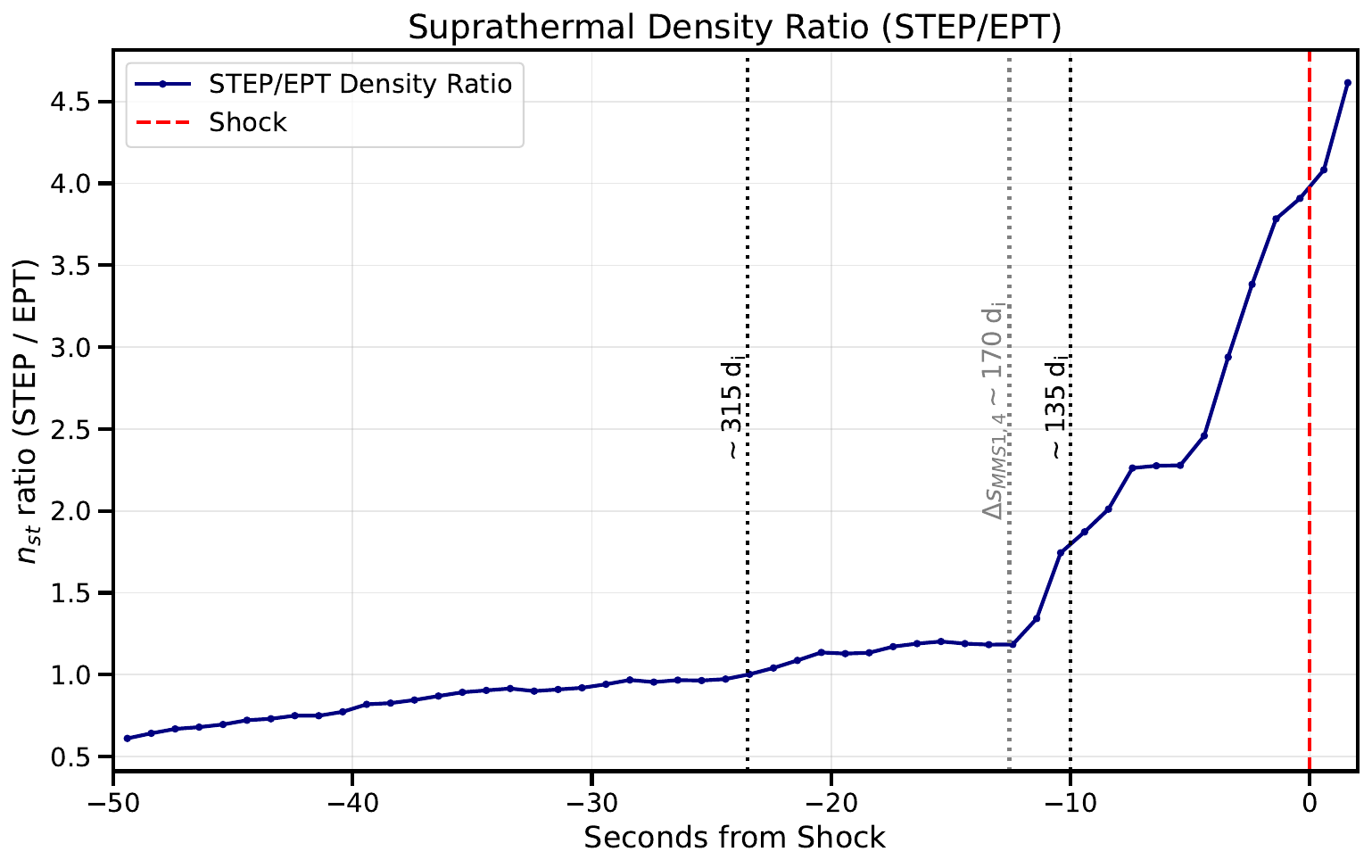}
    \caption{Suprathermal density ratio ($n_{st,STEP}/n_{st,EPT}$) observed by Solar Orbiter in the last 50 seconds upstream of the IP shock. The plot specifically highlights the transition where lower-energy suprathermals (STEP) overtake the energetic tail (EPT) in terms of number density. Vertical dotted lines mark specific distances from the shock in ion inertial lengths ($d_i$). The dominance crossover (Ratio $>1$) starts at $\sim 315~d_i$ (23 sec), while the steepest gradient in the STEP-to-EPT suprathermal ratio occurs at about $\sim 135~d_i$ (10 sec). The gray dotted line indicates the equivalent spatial separation of MMS4 and MMS1 ($\sim 170~d_i$) for comparison and to associate the observations fo Figures~\ref{fig:1} and \ref{fig:3}.}
    \label{fig:appendix_2}
\end{figure}

For Solar Orbiter, $n_{st}$ combines STEP and EPT. To perform this procedure, an average flux is obtained from the 15 STEP pixels, which is then combined with the flux from the sunward telescope of EPT. STEP data requires correction for dead-time saturation observed near the shock. We derive a time-varying calibration factor by cross-calibrating STEP against EPT in their overlapping energy range ($\sim$60--80~keV), as shown in Figure~\ref{fig:appendix_1}a. The combined spectrum (Figure~\ref{fig:appendix_1}d) is then integrated. As shown in Figure~\ref{fig:appendix_2}, the ratio of low-energy (STEP) to high-energy (EPT) flux reveals that suprathermals begin to dominate the density at $\sim 315~d_i$ (23~s) upstream, with the sharpest gradient occurring at $\sim 135~d_i$ (10~s).

At this point, we should also note that instrumental differences introduce opposite biases in the calculated suprathermal density ratio ($n_{st}/n_{sw}$). For MMS, the bulk density ($n_{sw}$) derived from FPI particle moments is typically an underestimate compared to spacecraft potential techniques. An underestimated denominator leads to an overestimation of the ratio. Conversely, Solar Orbiter utilizes the accurate spacecraft potential for $n_{sw}$, but its suprathermal sensors (STEP/EPT) have narrower fields of view compared to MMS. This limitation likely underestimates the suprathermal count ($n_{st}$), i.e., the numerator. Consequently, the Solar Orbiter suprathermal density ratio likely represents an underestimate, while the MMS ratio represents an overestimate.

\subsection{Shock and FCS Analysis and Spatial Transformation}

Shock analysis utilizes Minimum Variance Analysis (MVA) \citep{sonnerup1967magnetopause} to evaluate the polarization of FCSs, and magnetic and velocity coplanarity \citep{colburn1966discontinuities} to determine the shock normal and $\theta_{Bn}$. For the FCSs, MVA yields a boundary-normal coordinate system (LMN), defined such that $\mathbf{\hat{n}}$ corresponds to the minimum variance direction, while $\mathbf{\hat{l}}$ and $\mathbf{\hat{m}}$ denote the maximum and intermediate variance directions. We verified that the derived parameters for the shocks were robust against the choice of averaging windows \citep{Trotta2022b}; however, for the MMS shock normal calculation, while relatively well constrained and in agreement with model-derived normals \citep{mo2025comparison}, we observed that shifting the upstream and downstream intervals causes slight variations in the direction, a sensitivity consistent with recent findings \citep{toy2025automated}. Furthermore, the determined shock normal was predominantly aligned with the GSE X-axis. In this frame, the observed compressive structures possessed minimum variance directions ($\mathbf{\hat{n}}$) which are close to that of the shock's normal, while their dominant magnetic components lay in the coplanar and perpendicular planes. The shock speed $V_{sh}$ was calculated via mass flux conservation in the spacecraft frame for Solar Orbiter. For MMS, the ``string-of-pearls'' configuration precluded a proper tetrahedral timing analysis; nevertheless, timing estimates between MMS1, 2, and 3 were consistent with velocities derived via mass flux conservation, yielding values ranging from $10$ to $80$ km/s. To compare the propagating IP shock with the quasi-stationary bow shock, time-series data were converted to a spatial distance $S$ using the shock speed along shock's normal and normalized by the upstream ion inertial length $d_i$:
\begin{equation}
S = \frac{V_{\text{sh}} \cdot (t - t_0)}{d_i}
\end{equation}
where $t_0$ is the shock crossing time. Given the quasi-stationary nature of the bow shock, we utilized the calculated range of $V_{sh}$ to bound our spatial interpretation. Power spectral density (PSD) was computed using the Welch method \citep{welch2003use}. Finally, regarding compressive non-linear transients (FCS, SLAMS), we simply require that the wave amplitude reaches 1.5 times the background magnetic field magnitude ($B_{max}> 1.5 \cdot B_{bg}$) as indicated in Figure \ref{fig:3}.

\end{document}